\documentclass[letterpaper, 10 pt, conference]{ieeeconf}  
\makeatletter
\let\NAT@parse\undefined
\makeatother

\IEEEoverridecommandlockouts                              %

\usepackage{fancyhdr}

\usepackage{algorithm,algorithmic}
\usepackage{times}
\usepackage{cite}
\usepackage{multicol}
\usepackage[bookmarks=true]{hyperref}
\usepackage{amsmath}
\usepackage{bm}
\usepackage{amssymb}
\usepackage{graphicx}
\usepackage{multirow}
\usepackage{float}

\usepackage[caption=false, font=footnotesize]{subfig}

\usepackage{lipsum}
\pdfoutput=1
\usepackage[dvipsnames]{xcolor}
\usepackage{cancel}
\usepackage{stmaryrd}

\usepackage{calrsfs}
\DeclareMathAlphabet{\pazocal}{OMS}{zplm}{m}{n}

\newcommand{\calH}{\pazocal{H}}

\newcommand{\calP}{\pazocal{P}}

\newcommand{\calU}{\pazocal{U}}

\usepackage{eufrak}

\newcommand{\Rb}{\mathbb{R}}
\newcommand{\Sb}{\mathbb{S}}
\newcommand{\Zb}{\mathbb{Z}}

\newcommand{\vzero}{{\bf 0}}

\newcommand{\vGamma}{{\mathbf{\Gamma}}}

\newcommand{\rd}{{\mathrm d}}

\newcommand{\rT}{{\mathrm{T}}}

\newcommand{\vx}{{\bf x}}

\newcommand{\vI}{{\bf I}}

\newcommand{\vA}{{\bf A}}
\newcommand{\vR}{{\bf R}}

\newcommand{\vK}{{\bf K}}

\newcommand{\vF}{{\bf F}}

\newcommand{\vQ}{{\bf Q}}
\newcommand{\vB}{{\bf B}}
\newcommand{\vS}{{\bf S}}
\newcommand{\vX}{{\bf X}}

\newcommand{\vP}{{\bf P}}

\newcommand{\vPhi}{{\bf \Phi}}

\newcommand{\argmax}{\operatornamewithlimits{argmax}}
\newcommand{\argmin}{\operatornamewithlimits{argmin}}
\newcommand{\T}{^\mathrm{T}}

\newcommand{\bb}{{\bm b}}

\newcommand{\bd}{{\bm d}}
\newcommand{\be}{{\bm e}}

\newcommand{\bg}{{\bm g}}
\newcommand{\bh}{{\bm h}}

\newcommand{\bp}{{\bm p}}

\newcommand{\bu}{{\bm u}}

\newcommand{\bx}{{\bm x}}
\newcommand{\by}{{\bm y}}
\newcommand{\bz}{{\bm z}}

\newcommand{\bzeta}{{\bm \zeta}}
\newcommand{\blambda}{{\bm \lambda}}

\DeclareFontFamily{OT1}{pzc}{}
\DeclareFontShape{OT1}{pzc}{m}{it}{<-> s * [1.10] pzcmi7t}{}
\DeclareMathAlphabet{\mathpzc}{OT1}{pzc}{m}{it}

\usepackage{amsthm}
\usepackage{soul}

\newtheorem{problem}{Problem}

\title{\LARGE \bf
Nonlinear Robust Optimization for Planning and Control
}

\author{Arshiya Taj Abdul, Augustinos D. Saravanos and Evangelos A. Theodorou
\thanks{This work was supported by the ARO Award $\#$W911NF2010151. Augustinos Saravanos acknowledges financial support by the A. Onassis Foundation Scholarship.}
\thanks{Arshiya Taj Abdul and Augustinos D. Saravanos are with the School of Electrical and Computer Engineering, Georgia Institute of Technology, Atlanta, GA, 30332, USA.
        {\tt\footnotesize \{aabdul6, asaravanos\}@gatech.edu}}%
\thanks{Evangelos A. Theodorou is with the Daniel Guggenheim School of Aerospace Engineering, Georgia Institute of Technology, Atlanta, GA, 30332, USA.
        {\tt\footnotesize evangelos.theodorou@gatech.edu}}%
}

\begin{document}

\maketitle
\pagestyle{empty}

\begin{abstract}
This paper presents a novel robust trajectory optimization method for constrained nonlinear dynamical systems subject to unknown bounded disturbances. In particular, we seek optimal control policies that remain robustly feasible with respect to all possible realizations of the disturbances within prescribed uncertainty sets. To address this problem, we introduce a bi-level optimization algorithm. The outer level employs a trust-region successive convexification approach which relies on linearizing the nonlinear dynamics and robust constraints. The inner level involves solving the resulting linearized robust optimization problems, for which we derive tractable convex reformulations and present an Augmented Lagrangian method for efficiently solving them. To further enhance the robustness of our methodology on nonlinear systems, we also illustrate that potential linearization errors can be effectively modeled as unknown disturbances as well. Simulation results verify the applicability of our approach in controlling nonlinear systems in a robust manner under unknown disturbances. The promise of effectively handling approximation errors in such successive linearization schemes from a robust optimization perspective is also highlighted.
\end{abstract}

\section{Introduction}

Safety-critical trajectory optimization problems arise in a wide range of application domains, including autonomous driving \cite{howell2019altro, aoyama2024second}, unmanned aerial vehicles \cite{bonalli2019gusto, oleynikova2016continuous}, multi-agent systems \cite{saravanos2023_distributed_ddp, shorinwa2023distributed}, and many other fields. Such systems are often subject to complex nonlinear dynamics and underlying uncertainties, which pose major challenges in designing algorithms that are both robust and computationally efficient. Therefore, there is a great need for optimization frameworks that can effectively handle nonlinear dynamics, guarantee robust and safe operation under uncertainty, and maintain computational tractability.  

In most trajectory optimization approaches that explicitly address uncertainty, this is typically achieved through modeling as stochastic noise. Conventional approaches that belong in this category include LQG control \cite{todorov2005generalized, chen2021constrained, ma2022local}, covariance steering \cite{9029993, 8651541, liu2024optimal,  balci2022constrained, 10802104} and chance-constrained trajectory optimization algorithms \cite{9143595, 9861759, nakka2020chance, aoyama2021_receding}. 
Nevertheless, these methods can only guarantee safety and constraint satisfaction in a probabilistic sense.
This limitation makes them often impractical for safety-critical applications that require robustness guarantees under \textit{all} possible uncertainty realizations. 

Another approach for characterizing uncertainty is through unknown deterministic disturbances, modeled to lie inside prescribed bounded sets. This approach facilitates the development of trajectory optimization frameworks that guarantee safety and feasibility for all realizations of disturbances by optimizing for the worst-case scenario. This concept originated from the field of \textit{robust control} \cite{SAFONOV20121, ben2009robust}, which focuses on establishing stability and performance margins of control systems under parametric or exogenous uncertainties characterized as unknown deterministic disturbances \cite{green2012linear}. 

Despite the rising interest for achieving robust policies under uncertainty, relatively few works address the challenges of nonlinear dynamics and constraints. One class of related methods is min-max optimal control \cite{morimoto2002minimax, sun2018min}, where the uncertainty is characterized using an adversarial control policy modeled to lie inside a bounded set. 
However, such robust trajectory optimization methods fail to accommodate state constraints, which are crucial for establishing robustness in safety-critical applications. From a different point of view, robust model predictive control (MPC) typically focuses on linear systems \cite{richards2005robust, oravec2015alternative}. Tube-based methods are also widely used, yet their main disadvantage is conservatism due to decoupling the nominal control computation from the disturbance rejection \cite{mayne2011robust}. Consequently, there is a critical need for developing optimization frameworks that effectively address constrained nonlinear trajectory optimization problems under unknown disturbances.

Robust Optimization (RO) focuses on finding optimal solutions that remain robust feasible under all possible uncertainty realizations within some prescribed bounded sets \cite{ben2002robust, ben2009robust}. As such problems are inherently intractable due to the semi-infinite nature of the constraints, the main objective of RO techniques is to derive tractable reformulations or approximations. While there exists a rich amount of literature on applying RO methodologies on convex conic optimization problems \cite{ben2002robust, bertsimas2011theory}, extending these approaches to settings involving nonlinear/non-convex constraints remains a significant challenge \cite{leyffer2020survey}. This difficulty has naturally limited the applicability of RO in trajectory optimization, where nonlinear dynamics and constraints are prevalent.

%


The contribution of this paper is a novel robust trajectory optimization methodology which extends RO for constrained nonlinear dynamical systems under unknown disturbances. In particular, we present a bi-level optimization framework. The outer level consists of a sequential convexification scheme which linearizes the dynamics and constraints. At the inner level, we address the resulting linearized RO problems by deriving tractable convex reformulations and solving them using an Augmented Lagrangian (AL) method. Furthermore, we also highlight the ability of our framework to model potential linearization errors as disturbances for further enhancing its robustness on nonlinear systems. We showcase the efficacy and robustness of the proposed methodology through simulation experiments. We also analyze the impact of linearization error on the constraint violation highlighting the effectiveness of modeling through a RO perspective.

%

\textit{Organization of the Paper:} We begin by introducing the problem statement addressed in this work in Section \ref{Problem Statement}. Next, we present the successive linearization scheme in Section \ref{Successive Linearization Formulation}, followed by a methodology for solving the inner robust optimization problems in Section \ref{Inner Constrained Robust Optimization}. 
In Section \ref{Final Algorithm}, we provide the complete algorithm for our robust nonlinear trajectory optimization framework. We then extend the framework to incorporate linearization error as uncertainty in Section \ref{Linearization error as uncertainty}. The efficiency of the proposed frameworks is illustrated through simulation experiments in Section \ref{Simulation Results}. Section \ref{Conclusion} concludes our paper and provides future research directions.

\textit{Notations:} We represent the space of symmetric positive definite (semidefinite) matrices with dimension $n$ as $\Sb^{++}_n$ ($\Sb^{+}_n$). The 2-norm of a vector $\bx$ is denoted with $\| \bx \|_2$, while the Frobenius norm of a matrix $\vX$ is given by $\| \vX \|_F$. Further, a weighted 2-norm of a vector $\bx$ defined for a $\vQ \in \Sb^{+}_n$ as $\sqrt{\bx \T \vQ \bx}$ is denoted by $\| \bx \|_{\vQ}$. The indicator function of a set $X$, $\pazocal{I}_X$ is defined as $\pazocal{I}(\vx) = \{ 0$ if $\bx \in X$ or $+\infty$ if $\bx \notin X\}$. 
With $\llbracket a, b \rrbracket$, we denote the integer set $\{ [a,b] \cap \Zb \}$.

\section{Problem Statement} \label{Problem Statement}
Consider the following discrete-time nonlinear dynamics
\begin{align}
\bx_{k+1} & = f(\bx_k, \bu_k) + \bd_k, \quad k \in \llbracket 0, T-1 \rrbracket,
\label{original dynamics}
\\
\bx_0 & = \Bar{\bx}_0 + \Bar{\bd}_0,
\label{original dynamics init condition}
\end{align}
where $\bx_k \in \Rb^{n_x}$ is the state, $\bu_k \in \Rb^{n_u}$ is the control input, $f: \Rb^{n_x} \times \Rb^{n_u}  \rightarrow \Rb^{n_x}$ is the known dynamics function and $T$ is the time horizon. The terms $\bd_k \in \Rb^{n_x}$ represent \textit{unknown disturbances} which are formally defined below. In addition, the initial state $\bx_0$ consists of a known part $\bar{\bx}_0$, as well as an unknown part $\bar{\bd}_0$. 

In this work, we characterize the uncertainty to be lying in a bounded ellipsoidal set. This is quite common in most robust control applications, where ellipsoidal sets are used to model exogenous uncertainty \cite{Petersen2000}. Further, ellipsoidal sets can form a basis to address other more complex uncertainty sets \cite{BENTAL19991}. Considering $\bzeta = [\Bar{\bd}_0; \bd_0; \dots; \bd_{T-1} ]$, we define the uncertainty set
\begin{align*}
    \calU[\tau] = 
    \{ \bzeta | ~ \exists (\bz \in \Rb^{n_z}, \tau \in \Rb^{+}): ~
    \bzeta = \vGamma \bz, ~ \bz\T \vS \bz \leq \tau \},
\end{align*}
where $\bzeta \in \Rb^{(T+1) n_x}$, $\vGamma \in \Rb^{(T+1) n_x \times n_z}$, and $\vS \in \Sb^{++}_{n_z}$. Note that the positive definiteness of the matrix $\vS$ ensures that the uncertainty set is bounded. We also highlight that our methodology can be extended for other common types of uncertainty sets such as ellitopes, polytopes, etc. \cite{ben2009robust}.

Our system is also subject to the following \textit{robust} state and control constraints
%
%
\begin{align}
\bg(\bx; \bzeta) \leq 0, \quad \forall \bzeta \in \calU[\tau],
\\
\bh(\bu; \bzeta) \leq 0, \quad \forall \bzeta \in \calU[\tau],
\end{align}
where $\bx = [\bx_0; \bx_1; \dots; \bx_T]$, $\bu = [\bu_0; \bu_1; \dots; \bu_{T-1}]$, $\bg: \Rb^{(T+1)n_x} \rightarrow \Rb^{n_g}$ and $\bh: \Rb^{Tn_u} \rightarrow \Rb^{n_h}$. We emphasize that we seek for solutions that satisfy the above constraints for all possible realizations of the disturbances $\bzeta$ within the uncertainty set $\calU[\tau]$. In particular, we seek affine control policies of the following form
\begin{align}
    \bu_k = \Bar{\bu}_k + \vK_k \bd_{k-1} , 
    \label{control input expression}
\end{align}
where $\Bar{\bu}_k \in \Rb^{n_u}$ are feed-forward controls and $\vK_k \in \Rb^{n_u \times n_x}$ are feedback gains. By convention, we set $\bd_{-1} = \bar{\bd}_0$. We will now introduce the robust trajectory optimization problem addressed in this work. 

\begin{problem}[Robust Trajectory Optimization Problem] Find the optimal control policy $\{ \bar{\bu}_k, \vK_k \}_{k=0}^{T-1}$ such that
\begin{equation}
\begin{aligned}
    & \min_{\{\Bar{\bu}_k, \vK_k \}_{k=0}^{T-1}} 
    \sum_{k=0}^{T-1} \Bar{\bu}_k^\rT \vR_u^k \Bar{\bu}_k
    + \| \vR_K^k \vK_k \|_F^2
    \\
    \text{s.t. } \quad
    & 
     \bx_{k+1} = f(\bx_k, \bu_k) + \bd_k,\quad 
     \bx_0 = \Bar{\bx}_0 + \Bar{\bd}_0, 
    \\
    & \bg (\bx; \bzeta) \leq 0, \quad
    \bh (\bu; \bzeta) \leq 0, 
    \quad \forall \bzeta \in \calU[\tau].
\end{aligned}
\label{Original Problem}
\end{equation}
\label{problem: original problem}
\end{problem}

The above problem is especially challenging because of the following inherent difficulties. First, Problem \ref{problem: original problem} is not tractable at its current form as it is subject to an infinite amount of constraints. The robust constraints are the constraints that need to be satisfied for all possible realizations of the uncertainty $\bzeta$ lying in the defined uncertainty set $\calU[\tau]$. The second difficulty stems from the nonlinearity in the dynamics and the constraints which do not allow for the direct application of standard robust optimization (RO) techniques to convert the problem into a tractable form. The following sections present a methodology that integrates robust optimization, successive linearization and operator splitting through the Alternating Direction Method of Multipliers \cite{boyd2011distributed} towards effectively addressing these challenges. 
\section{Successive Linearization Approach}
\label{Successive Linearization Formulation}
We start with introducing a successive linearization scheme for addressing the nonlinearities in Problem \ref{problem: original problem}. 
Let us define the disturbance-free states $ \Bar{\bx}_k$ whose dynamics are obtained through simply setting $\bzeta = \vzero$, i.e.,
\begin{equation}
    \bar{\bx}_{k+1} = f(\bar{\bx}_k, \bar{\bu}_k), \quad k \in \llbracket 0, T-1 \rrbracket.
\end{equation}
Then, the disturbance component $\bx_k^\rd$ of the state is given by
\begin{equation}
    \bx_k^\rd = \bx_k - \bar{\bx}_k.
    \label{x_d expression}
\end{equation}

\subsubsection{Dynamics Linearization}
To construct a successive linearization scheme, we consider linearizing the dynamics \eqref{original dynamics}, around a nominal trajectory $\{ \hat{\bx}_k, \hat{\bu}_k \}_{k=0}^{T-1}$, whose dynamics are given by 
\begin{align}
\hat{\bx}_{k+1} & = f(\hat{\bx}_k, \hat{\bu}_k), \quad k \in \llbracket 0, T-1 \rrbracket,
\label{nominal dynamics}
\\
\hat{\bx}_0 & = \bar{\bx}_0.
\label{nominal dynamics init condition}
\end{align}
Thus, we obtain the following linearized dynamics 
\begin{equation}
\bx_{k+1} - \hat{\bx}_{k+1}
=  \vA_k ( \bx_{k} -  \hat{\bx}_{k})
+ \vB_k ( \bu_k - \hat{\bu}_k )
+ \bd_k
\label{linearized dynamics - ver0}
\end{equation}
with $\vA_k \in \Rb^{n_x \times n_x}$ and $\vB_k \in \Rb^{n_x \times n_u}$ given as
\begin{equation*}
\vA_k = \nabla_{\bx} f(\bx_k, \bu_k) \big|_{\substack{\bx_k = \hat{\bx}_k \\ \bu_k = \hat{\bu}_k}},
\quad
\vB_k = \nabla_{\bu} f(\bx_k, \bu_k) \big|_{\substack{\bx_k = \hat{\bx}_k \\ \bu_k = \hat{\bu}_k}}.
\end{equation*}
Next, we define the deviations $\delta \hat{\bx}_{k} = \bar{\bx}_{k} - \hat{\bx}_{k}$, $\delta \hat{\bu}_k = \bar{\bu}_{k} - \hat{\bu}_{k}$, such that using \eqref{control input expression}, \eqref{x_d expression}, we can rewrite \eqref{linearized dynamics - ver0} as
\begin{equation}
\begin{aligned}
    \delta \hat{\bx}_{k+1} + \bx_{k+1}^\rd
    & =  \vA_k (\delta \hat{\bx}_{k} + \bx_{k}^\rd)
    \\
    &~~~~~ + \vB_k ( \delta \hat{\bu}_k + \vK_k \bd_{k-1} )
    + \bd_k
\end{aligned}
\label{linearized dynamics - ver1}
\end{equation}
or in a more compact form as
\begin{equation}
\delta \hat{\bx} + \bx^\rd
= 
\vF_0 \delta \hat{\bx}_0
+ \vF_u ( \delta \hat{\bu} + \vK \bzeta )
+ \vF_0 \bx^\rd_{0}
+ \vF_{d} \bzeta
\label{linearized dynamics - ver1p5}
\end{equation}
where
$\bx^\rd = [\bx^\rd_0; \dots; \bx^\rd_{T}]$, 
$\delta \hat{\bx} = [\delta \hat{\bx}_0; \dots; \delta \hat{\bx}_{T}]$, 
$\delta \hat{\bu} = [\delta \hat{\bu}_0; \dots; \delta \hat{\bu}_{T-1}]$, $\vK = \mathrm{bdiag} (\{ \vK_k \}_{k=0}^{T-1})$ and the matrices $\vF_0$, $\vF_u$, and $\vF_{d}$ are defined in Appendix \ref{Linearized Dynamics matrices}. Finally, given that $\delta \hat{\bx}_0 = 0$ and $\bx^\rd_{0} = \bar{\bd}_0$, we can further simplify \eqref{linearized dynamics - ver1p5} to
\begin{equation}
\delta \hat{\bx} + \bx^\rd
= 
\vF_u ( \delta \hat{\bu} + \vK \bzeta )
+ \vF_{\zeta} \bzeta
\label{linearized dynamics - ver2}
\end{equation}
where the matrix $\vF_{\zeta}$ is also defined in Appendix \ref{Linearized Dynamics matrices}. 

\subsubsection{Robust Constraints Linearization}
Subsequently, we also linearize the robust constraints. Note that linear constraints, e.g. control box constraints, can be directly incorporated. Without loss of generality, we limit our exposition to the state constraints $\bg (\bx; \bzeta) \leq 0$. In particular, we obtain the linearized robust constraint
\begin{equation}
    \bg(\hat{\bx}) + \nabla_x \bg(\hat{\bx}) ( \delta \hat{\bx} + \bx^\rd ) \leq 0, \quad \forall \bzeta \in \calU[\tau],
\label{Linearized constraint - ver1}
\end{equation}
around the nominal trajectory $\{ \hat{\bx}_k, \hat{\bu}_k \}_{k=0}^{T-1}$. By combining the linearized dynamics \eqref{linearized dynamics - ver1} with \eqref{Linearized constraint - ver1}, we arrive to the equivalent constraint
\begin{equation}
    \bg^{\text{lin}}(\delta \hat{\bu}, \vK; \bzeta)
    \leq 0,
    \quad
    \forall \bzeta \in \calU[\tau],
\label{linearized constraint final}
\end{equation}
where
\begin{equation}
    \bg^{\text{lin}} = \bg(\hat{\bx}) + \nabla_x \bg(\hat{\bx}) \big( 
    \vF_u ( \delta \hat{\bu} + \vK \bzeta )
    + \vF_{\zeta} \bzeta).
\end{equation}
%
%

\subsubsection{Successive Linearization Scheme}
Based on the previous linearizations, we propose an iterative scheme which solves the following problem in place of Problem \ref{problem: original problem}.
\begin{problem}[Linearized Problem] Find the optimal decision variables $\delta \hat{\bu}, \vK$ such that
\begin{subequations}
\begin{align}
    \min_{\delta \hat{\bu}, \vK} \; \;
    & \pazocal{Q}_{\hat{u}} (\delta \hat{\bu})
    + \pazocal{Q}_K (\vK) \nonumber
    \\
    \mathrm{s.t.} \quad
    & \bg^{\text{lin}}(\delta \hat{\bu}, \vK; \bzeta)
    \leq 0,
    \quad
    \forall \bzeta \in \calU[\tau],
    \\
    & 
    \| \vF_u \delta \hat{\bu} \|_2 \leq r_{\text{trust}},
    \label{trust region constraint}
\end{align}%
\label{Linearized Problem}%
\end{subequations}
where $\pazocal{Q}_{\hat{u}} (\delta \hat{\bu}) = \sum_{k=0}^{T-1} ( \hat{\bu}_k + \delta \hat{\bu}_k )\T \vR_u^k ( \hat{\bu}_k + \delta \hat{\bu}_k)$, $\pazocal{Q}_K (\vK) = \sum_{k=0}^{T-1} \| \vR_K^k \vK_k \|_F^2$.
\label{problem: linearized problem}
\end{problem}
%
%
%
The trust region constraint \eqref{trust region constraint} with $r_{\text{trust}} \in \Rb^{+}$ is added to ensure the boundedness of the linearized problem. The solution of Problem \ref{problem: linearized problem} would provide a new nominal trajectory $\{ \hat{\bx}_k, \hat{\bu}_k \}_{k=0}^{T-1}$ around which we perform a new linearization, and so on - as described in the next section. Nevertheless, we still cannot solve the above problem due to two prominent issues. The first is computational intractability due to the constraint \eqref{linearized constraint final}, while the second one is temporary infeasibilities that might arise in the linearized problems.
%
%

%
\section{Inner Constrained Robust Optimization}
\label{Inner Constrained Robust Optimization}
\subsection{Tractable Reformulation of Robust Optimization Problem}
\label{Tractable Reformulation of Robust Optimization Problem}
In this section, we transform Problem \ref{problem: linearized problem} into tractable form. We start by splitting the constraint \eqref{linearized constraint final} through introducing a slack variable $\bp = [p_1; \dots; p_{n_g}] \in \Rb^{n_g}$, which leads to the following set of constraints
\begin{align}
    & \bg^{\text{lin},1}(\delta \hat{\bu}, \bp) := \bg(\hat{\bx}) + \nabla_x \bg(\hat{\bx}) 
    \vF_u \delta \hat{\bu} 
    + \bp \leq 0,
    \label{linearized const - without dist split}
   \\
    & \bg^{\text{lin},2}(\vK, \bp; \bzeta) := \nabla_x \bg(\hat{\bx}) (  \vF_u \vK + \vF_{\zeta}) \bzeta 
    \nonumber
    \\
    & ~~~~~~~~~~~~~~~~~~~~~~~~~~~~~~~~~~~~~~~~~ \leq \bp,
    ~ 
    \forall \bzeta \in \calU[\tau].
    \label{linearized const - dist split}
\end{align}
Note that only the constraint \eqref{linearized const - dist split} now depends on the uncertainty $\bzeta$, and thus remains intractable. We transform it into a tractable form by first reformulating it as 
\begin{equation}
    \max_{\bzeta \in \calU[\tau]} \nabla_x \bg(\hat{\bx}) (  \vF_u \vK + \vF_{\zeta}) \bzeta \leq \bp,
    \label{max constraint formulation}
\end{equation}
which implies that the constraint needs to be satisfied for the worst-case scenario. 
Next, by using the first-order optimality conditions \cite{abdul2024scaling}, we obtain for each $j \in \llbracket 1, n_g \rrbracket$,
\begin{equation}
\begin{aligned}
     & \max_{\bzeta \in \calU[\tau]} \nabla_x \bg_j(\hat{\bx})^\rT (  \vF_u \vK + \vF_{\zeta}) \bzeta 
     \\
    &~~~~~~
    = (\tau)^{1/2} \| \vGamma^\rT ( \vF_u \vK + \vF_{\zeta} )^\rT \nabla_x \bg_j(\hat{\bx}) \|_{\vS^{-1}},
\end{aligned}
\end{equation}
using which, the constraint \eqref{max constraint formulation} can be equivalently given by the following set of tractable constraints 
\begin{equation}
    g_j^{\text{trac}}(\vK, \bp) \leq 0, 
    \; j \in \llbracket 1, n_g \rrbracket,
    \label{linearized tractable constraint}
\end{equation}
with
\begin{equation*}
    g_j^{\text{trac}}(\vK, \bp) := (\tau)^{1/2} \| \vGamma^\rT ( \vF_u \vK + \vF_{\zeta} )^\rT \nabla_x \bg_j(\hat{\bx}) \|_{\vS^{-1}} - p_j.
\end{equation*}
Thus, Problem \ref{problem: linearized problem} is equivalent to the following convex tractable reformulation.
\begin{problem}[Tractable Linearized Problem]
\label{Tractable Linearized Problem}
Find the optimal decision variables $\delta \hat{\bu}, \vK, \bp$ such that
\begin{equation*}
\begin{aligned}
    \min_{\delta \hat{\bu}, \vK, \bp}\; \;
    & \pazocal{Q}_{\hat{u}} (\delta \hat{\bu})
    + \pazocal{Q}_K (\vK) 
    \\
    \text{s.t.} \quad 
    & \bg^{\text{lin},1}(\delta \hat{\bu}, \bp) \leq 0,
    \\
    & g_j^{\text{trac}}(\vK, \bp) \leq 0, 
    \; j \in \llbracket 1, n_g \rrbracket,
    \\
    & \| \vF_u \delta \hat{\bu} \|_2 \leq r_{\text{trust}}.
\end{aligned}
\label{Linearized Problem - tractable version}
\end{equation*}
\end{problem}
%
%
%
%
\subsection{ADMM for Solving Constrained Robust Optimization}
\label{ADMM for Solving Constrained Robust Optimization}
Subsequently, we present a method for solving Problem \ref{Tractable Linearized Problem}, based on the Alternating Direction Method of Multipliers (ADMM). ADMM is an AL approach for solving convex optimization problems of the form 
\begin{equation}
    \min_{\by \in Y, \Tilde{\by} \in \tilde{Y}} \; \calH_y(\by) + \calH_{\tilde{y}} (\Tilde{\by})
    \quad
   \text{s.t. } \vP \by + \vQ \Tilde{\by} = \bb,
   \label{ADMM problem form}
\end{equation}
where $\calH_y$ and  $\calH_{\tilde{y}}$ are closed, proper, and convex functions \cite{boyd2011distributed}. This framework solves the problem in a distributed manner with respect to the variables $\by$ and $\Tilde{\by}$ (typically referred to as ADMM blocks). However, we are using this algorithm due to its key feature of infeasibility detection \cite{banjac2019infeasibility, 7040300}. 
Under certain mild assumptions, if the cost function $\calH_y$ and  $\calH_{\tilde{y}}$ are convex quadratic and the matrix $[\vP, \vQ]$ has full rank, then the ADMM converges to a solution, lying in the set $\{ (\by, \tilde{\by}): \by \in Y, \Tilde{\by} \in \tilde{Y} \}$, that has the minimum Euclidean distance to the set $\{ (\by, \tilde{\by}): \vP \by + \vQ \Tilde{\by} = \bb \}$ \cite{7040300}. Thus, we can employ ADMM to inexactly solve Problem \ref{Tractable Linearized Problem}.

To achieve this, we need to first transform Problem \ref{Tractable Linearized Problem} to the form \eqref{ADMM problem form}. For that, we would rewrite the set of constraints \eqref{linearized tractable constraint} using the slack variable $\tilde{\bp} \in \Rb^{n_g}$ as the following equivalent constraints
\begin{align}
    & 
    g_j^{\text{trac}}(\vK, \tilde{\bp}) \leq 0, \quad j \in \llbracket 1, n_g \rrbracket, 
    \label{admm constraint 2}
    \\
    & 
    \bp = \tilde{\bp}.
    \label{admm constraint 3}
\end{align}
%
Thus, Problem \ref{Tractable Linearized Problem} can be equivalently expressed in the following form.
\begin{problem}[Tractable Linearized Problem - ADMM form]
\label{Tractable Linearized Problem - ADMM}
Find the optimal decision variables $\delta \hat{\bu}, \vK, \bp, \tilde{\bp}$ such that
\begin{equation*}
\begin{aligned}
   & \min_{\delta \hat{\bu}, \vK, \bp, \tilde{\bp}} \; \;
    \pazocal{H}_{p} ( \delta \hat{\bu}, \bp )
    + \pazocal{H}_{\tilde{p}} ( \vK, \tilde{\bp}  )
    \\
    &~~~ \qquad
    \text{s.t.} \quad \bp = \tilde{\bp}
\end{aligned}
\label{Linearized Problem - admm version}
\end{equation*}
with $\pazocal{H}_{p} ( \delta \hat{\bu}, \bp ) = \pazocal{I}_{\calP} + \pazocal{Q}_{\hat{u}} (\delta \hat{\bu})$, 
$\pazocal{H}_{\tilde{p}} ( \vK, \tilde{\bp}  ) = \pazocal{I}_{\tilde{\calP}} +  \pazocal{Q}_K (\vK)$, 
where $\calP$ represents the feasible region for the set of constraints \eqref{trust region constraint}, \eqref{linearized const - without dist split}, and $\tilde{\calP}$ represents the feasible region for the constraint \eqref{admm constraint 2}.
\end{problem}
%

To solve Problem \ref{Tractable Linearized Problem - ADMM} using ADMM, we start with formulating the AL as follows
\begin{equation}
\begin{aligned}
    \pazocal{L}_{\rho} ( \delta \hat{\bu}, \bp, \vK, \tilde{\bp}; \blambda )
    & = \pazocal{H}_{p} ( \delta \hat{\bu}, \bp )
    + \pazocal{H}_{\tilde{p}} ( \vK, \tilde{\bp}  )
    \\
    &~~~
    + \blambda\T (\bp - \tilde{\bp})
    + \frac{\rho}{2} \| \bp - \tilde{\bp} \|_2^2,
\end{aligned}
\end{equation}
where $\blambda \in \Rb^{n_g}$ is the dual variable for the constraint $\bp = \tilde{\bp}$ and $\rho >0$ is the penalty parameter. We consider the variables $\{ \vK, \tilde{\bp} \}$ as the first block, and $\{ \delta \hat{\bu}, \bp \}$ as the second block of ADMM. Each ADMM iteration - indexed with $l_{\text{in}}$ - would then involve the following sequential steps:
\begin{align}
    \text{i) } & \{ \vK, \tilde{\bp} \}^{l_{\text{in}}} = \argmin_{\vK, \tilde{\bp}} \pazocal{L}_{\rho} (\{ \delta \hat{\bu}, \bp\}^{l_{\text{in}}-1}, \vK, \tilde{\bp}; \blambda^{l_{\text{in}}-1} ) \nonumber
    \\
    \text{ii) } &
    \{ \delta \hat{\bu}, \bp \}^{l_{\text{in}}}= 
    \argmin_{\delta \hat{\bu}, \bp} \pazocal{L}_{\rho} ( \delta \hat{\bu}, \bp,\{ \vK, \tilde{\bp} \}^{l_{\text{in}}};\blambda^{l_{\text{in}}-1} )
    \nonumber
    \\
    \text{iii) } &
    \blambda^{l_{\text{in}}} = \argmax_{\blambda} \pazocal{L}_{\rho} ( \{ \delta \hat{\bu}, \bp, \vK, \tilde{\bp} \}^{l_{\text{in}}};\blambda )
    \nonumber
\end{align}
%
The above updates can finally be rewritten as follows 
\begin{subequations}
\begin{align}
    & \{ \vK, \tilde{\bp} \}^{l_{\text{in}}} 
    =
    \argmin_{ \vK, \tilde{\bp} } 
    \pazocal{Q}_K (\vK)
    + \blambda^{l_{\text{in}}-1}{}^\rT (\bp^{l_{\text{in}}-1} - \tilde{\bp})
    \nonumber \\
    & \qquad\qquad\qquad\qquad\qquad\qquad + \frac{\rho}{2} \| \bp^{l_{\text{in}}-1} - \tilde{\bp} \|_2^2
    \label{ADMM - block 1 update}
    \\[0.2cm]
    & \qquad\qquad \text{s.t.} \quad g_j^{\text{trac}}(\vK, \tilde{\bp}) \leq 0, \quad j \in \llbracket 1, n_g \rrbracket,  \nonumber
    \\[0.2cm]
    & \{ \delta \hat{\bu}, \bp \}^{l_{\text{in}}} 
    =
    \argmin_{\delta \hat{\bu}, \bp } \pazocal{Q}_{\hat{u}} (\delta \hat{\bu})
    + \blambda^{l_{\text{in}}-1}{}^\rT (\bp - \tilde{\bp}^{l_{\text{in}}})
    \nonumber \\
    & \qquad\qquad\qquad\qquad\qquad\qquad
    + \frac{\rho}{2} \| \bp - \tilde{\bp}^{l_{\text{in}}} \|_2^2
    \label{ADMM - block 2 update}
     \\[0.2cm]
    & \qquad \qquad
    \text{s.t. } \quad \bg^{\text{lin},1}(\delta \hat{\bu}, \bp) \leq 0, ~ \| \vF_u \delta \hat{\bu} \|_2 \leq r_{\text{trust}}, \nonumber
    \\[0.2cm]
    & \lambda^{l_{\text{in}}} =
    \lambda^{l_{\text{in}}-1}
    + \rho (\bp^{l_{\text{in}}} - \tilde{\bp}^{l_{\text{in}}}).
    \label{ADMM - dual update}
\end{align}
\end{subequations}
\section{Final Algorithm}
\label{Final Algorithm}
 \begin{algorithm}[t]
 \caption{Nonlinear Robust Trajectory Optimization}
 \begin{algorithmic}[1]
 \renewcommand{\algorithmicrequire}{\textbf{Input:}}
 \renewcommand{\algorithmicensure}{\textbf{Output:}}
 \REQUIRE Uncertainty set parameters $\vS$, $\tau$, $\bar{\bx}_0$
 \\ \textit{Initialization}: $\hat{\bu}$, $\rho$, $\blambda$, $\bp, \tilde{\bp}$, $r_{\text{trust}}$, $\alpha \in [0.5,1)$, $\beta>1$, $\eta_1, \eta_2, r_{\text{min}}, \rho_{\text{max}} \in \Rb^{+}$  
  \WHILE{\textit{not converged}}
  \STATE \textit{Set the nominal trajectory $(\hat{\bx}, \hat{\bu})$ :}
  \STATE
  $\hat{\bx}_0 = \bar{\bx}_0$,
  \FOR  {$k = 0$ to $T-1$}
  \STATE $\hat{\bx}_{k+1} = f(\hat{\bx}_{k}, \hat{\bu}_k)$ 
  \ENDFOR
  \STATE\textit{Solve the Problem \ref{Linearized Problem - tractable version} linearized around $(\hat{\bx}, \hat{\bu})$:}
  \IF{$ \| \bp - \tilde{\bp} \|_2 \leq \epsilon_p$}
    \STATE Solve Problem \ref{Linearized Problem - tractable version}  $\leftarrow \delta \hat{\bu}$, $\vK$.
    \STATE $\bp = \tilde{\bp}$
    \ELSE
     \STATE 
     Inexactly solve Problem \ref{Tractable Linearized Problem - ADMM} using the Algorithm \ref{alg2}
    $\leftarrow \delta \hat{\bu}$, $\vK$, $\blambda$, $\bp $, $ \tilde{\bp}$
  \ENDIF
  \STATE
  $\hat{\bu} = \hat{\bu} + \delta \hat{\bu}$
  \IF{$\| \delta \hat{\bu}\|_2 \leq \epsilon_{\hat{u}}$ 
  and 
  $\bp = \tilde{\bp}$}
  \STATE break
  \ENDIF
  \IF{ $\| \delta \hat{\bu} \|_2 \geq \eta_1 \| \bp - \tilde{\bp} \|_2$} 
    \STATE reduce $r_{\text{trust}}$: $r_{\text{trust}} = \min( \alpha r_{\text{trust}}, r_{\text{min}})$ 
  \ELSIF{  $\| \delta \hat{\bu}\|_2 \leq \eta_2 \| \bp - \tilde{\bp} \|_2$   }
  \STATE Increase $\rho$: $\rho = \max( \beta \rho, \rho_{\text{max}} ) $
  \ENDIF
  \ENDWHILE
 \ENSURE  $\hat{\bu}, \vK$
 \end{algorithmic}
 \label{alg1}
 \end{algorithm}
 \vspace{-0.2cm}
\begin{algorithm}[t]
 \caption{Inner ADMM Loop}
 \begin{algorithmic}[1]
 \renewcommand{\algorithmicrequire}{\textbf{Input:}}
 \renewcommand{\algorithmicensure}{\textbf{Output:}}
 \REQUIRE $\blambda$, $\bp $, $ \tilde{\bp}$, $\rho$
 \STATE \textit{Initialize } $\blambda^0 = \blambda$, $\bp^0 = \bp$, $\tilde{\bp}^0 = \tilde{\bp}$ 
  \FOR {$l_{\text{in}} = 1$ to $L_{\text{max}}^{\text{in}}$}
    \STATE Sequentially solve \eqref{ADMM - block 1 update}, \eqref{ADMM - block 2 update}, and \eqref{ADMM - dual update} 
    \IF{$\| \bp^{l_{\text{in}}} - \tilde{\bp}^{l_{\text{in}}} \|_2 \leq \epsilon_p$}
    \STATE break
    \ENDIF
  \ENDFOR
  \STATE 
  $\delta \hat{\bu} = \delta \hat{\bu}^{l_{\text{in}}}$,
  $\vK = \vK^{l_{\text{in}}}$
  \STATE
  $\bp = \bp^{l_{\text{in}}}$,
  $\tilde{\bp} = \tilde{\bp}^{l_{\text{in}}}$,
  $\blambda = \blambda^{l_{\text{in}}}$
 \ENSURE  $\delta \hat{\bu}, \vK, \bp, \tilde{\bp}, \blambda$
 \end{algorithmic}
 \label{alg2}
 \end{algorithm}
\vspace{0.15cm}
In this section, we combine the above techniques to present the complete version of the bi-level optimization framework in Algorithm \ref{alg1}. First, the nominal control $\hat{\bu}$, penalty parameter $\rho$, dual variable $\blambda$, and slack variables $\bp$ and $\tilde{\bp}$ need to be initialized. In each outer iteration, we linearize Problem \ref{problem: original problem} around the nominal trajectory $\{ \hat{\bx}_k, \hat{\bu}_k \}_{k=0}^{T-1}$ to obtain Problem \ref{Linearized Problem - tractable version}. 
Nevertheless, as discussed earlier, Problem \ref{Linearized Problem - tractable version} can be infeasible. 
Thus, for the initial outer iterations, Problem \ref{Tractable Linearized Problem - ADMM} (an equivalent version of Problem \ref{Tractable Linearized Problem}) is inexactly solved using ADMM as disclosed in Algorithm \ref{alg2}. 
This involves sequentially solving \eqref{ADMM - block 1 update}, \eqref{ADMM - block 2 update}, and \eqref{ADMM - dual update} for a set number of ADMM iterations $L_{\text{max}}^{\text{in}}$. This approach is continued until the residual $\| \bp - \tilde{\bp}\|_2$ falls below a set threshold $\epsilon_p>0$. Once this happens, in each outer iteration, Problem \ref{Linearized Problem - tractable version} is directly solved to obtain $\delta \hat{\bu}$, $\vK$. 
Further, in each outer iteration, the trust region radius $r_{\text{trust}}$ and the penalty parameter $\rho$ are updated using the parameters $\alpha$ and $\beta$ respectively, based on the residuals $\| \delta \hat{\bu}\|_2$ and $\| \bp - \tilde{\bp}\|_2$. The algorithm is terminated when the following convergence criteria are fulfilled 
\begin{equation}
\| \delta \hat{\bu}\|_2 \leq \epsilon_{\hat{u}} \; 
\text{and}
\;
\bp = \tilde{\bp}
\end{equation}
where $\epsilon_{\hat{u}}>0$ is a set threshold.
Note that it is possible that the above framework would still not provide a completely robust solution for the actual nonlinear system as it relies on linearization techniques which might contain approximation errors. In the subsequent section, we address this challenge by presenting an extension of this framework which also models the linearization errors through a RO point of view. 
%
%
\section{Linearization error as uncertainty}
\label{Linearization error as uncertainty}
In this section, we present an extension of the above proposed framework to address the approximation errors that arise in successive linearization schemes. Note that Algorithm \ref{alg1} ensures that a \textit{linearized} trajectory given as 
\begin{align}
    \bx^{\text{lin}} = \bar{\bx} + \bx^d = 
    \bar{\bx} 
    + (\vF_u \vK + \vF_{\zeta}) \bzeta 
\end{align}
satisfies all the constraints in the robust sense. However, this might not ensure that the actual trajectory $\bx$ would satisfy all the constraints due to the linearization error ( i.e., $\bx - \bx^{\text{lin}}$). This reveals the need for ensuring robustness even in the presence of such linearization errors. For that, we interpret the linearization error as an additional source of uncertainty, such that we have the following instead of \eqref{linearized dynamics - ver2}
\begin{equation}
\delta \hat{\bx} + \bx^\rd
    = 
    \vF_u ( \delta \hat{\bu} + \vK \bzeta ) + \bx^e
\end{equation}
where $\bx^e \in \Rb^{(T+1)n_x}$ is the linearization error defined as $\bx^e = [\bx^e_0; \bx^e_1; \dots; \bx^e_{T}]$, with each of its components $\bx^e_k$ lying in the following ellipsoidal uncertainty sets
\begin{equation}
\begin{aligned}
    \pazocal{E}^k [\tau^{e}_k] 
    & = \{ \bx^e_k \in \Rb^{n_x} | \;
    \exists (\be^c_k \in \Rb^{n_x}, \tau^{e}_k \in \Rb^{+}) \; :  
    \\
    &~~~~~~~~~~~
   (\bx^e_k - \be^c_k)\T \vS^e_k (\bx^e_k - \be^c_k) \leq \tau^{e}_k \},
\end{aligned}
\end{equation}
where $\vS^e_k \in \Sb^{++}_{n_x}$.
We incorporate this modification into Algorithm \ref{alg1} by replacing the set of constraints \eqref{admm constraint 2} with
\begin{equation}
\begin{aligned}
    & 
    g_j^{\text{trac}}(\vK, \tilde{\bp}) 
    + \nabla_x \bg_j(\hat{\bx})\T \bx^e  
    \leq 0
    ~~
    \forall \{ \bx^e_k \in \pazocal{E}^k \}_{k=0}^{T}.
\end{aligned}
\label{lin const - error approach}
\end{equation}
Then, we have
\begin{equation}
    \nabla_x \bg_j(\hat{\bx})\T \bx^e
    = \sum_{k = 0}^{T} \nabla_{x_k} \bg_{j}(\hat{\bx})\T \bx^e_k,
\end{equation}
using which, we can rewrite the constraint \eqref{lin const - error approach} as follows 
\begin{equation}
\begin{aligned}
    & 
    g_j^{\text{trac}}(\vK, \tilde{\bp})
    + \sum_{k=0}^{T} \max_{ \bx^e_k \in \pazocal{E}^k } 
    \nabla_{x_k} \bg_{j}(\hat{\bx})\T \bx^e_k
    \leq 0.
\end{aligned}
\label{lin const - error eq3}
\end{equation}
Let us now simplify $\max_{ \bx^e_k \in \pazocal{E}^k } \nabla_{x_k} \bg_{j}(\hat{\bx})\T \bx^e_k$ by defining a variable 
$\tilde{\bx}_k^e = \bx^e_k - \be^c_k$ such that we obtain
\begin{equation}
\begin{aligned}
    & \max_{ \bx^e_k \in \pazocal{E}^k } \nabla_{x_k} \bg_{j}(\hat{\bx})\T \bx^e_k
    \\
    &~~
    =
    \nabla_{x_k} \bg_{j}(\hat{\bx})\T  \be^c_k
    + 
    \max_{ \tilde{\bx}_k^e{}\T \vS^e_k \tilde{\bx}_k^e \leq \tau^{e}_k  }
    \nabla_{x_k} \bg_{j}(\hat{\bx})\T  \tilde{\bx}_k^e.
\end{aligned}
\label{lin const - error eq1}
\end{equation}
By using the first order optimality conditions, we get 
\begin{equation}
\begin{aligned}
    \max_{  \tilde{\bx}_k^e{}\T \vS^e_k \tilde{\bx}_k^e \leq \tau^{e}_k   }
    & \nabla_{x_k} \bg_{j}(\hat{\bx})\T  \tilde{\bx}_k^e
    =
    ( \tau^{e}_k )^{1/2} \| \nabla_{x_k} \bg_{j}(\hat{\bx}) \|_{(\vS^e_k)^{-1}}
\end{aligned}
\label{lin const - error eq2}
\end{equation}
Using \eqref{lin const - error eq1} and \eqref{lin const - error eq2}, we can then rewrite \eqref{lin const - error eq3} as follows 
\begin{equation}
\begin{aligned}
    & 
    g_j^{\text{trac}}(\vK, \tilde{\bp})
    + 
    g_j^{\text{err}} 
    \leq 0
\end{aligned}
\label{modified admm constraints 2}
\end{equation}
where 
\begin{equation*}
\begin{aligned}
     g_j^{\text{err}}   & =
    \sum_{k=0}^{T} \bigg[ \nabla_{x_k} \bg_{j}(\hat{\bx})\T \be^c_k
    + ( \tau^{e}_k )^{1/2} \| \nabla_{x_k} \bg_{j}(\hat{\bx}) \|_{(\vS^e_k)^{-1}} \bigg].
\end{aligned}
\end{equation*}


The extension of Algorithm \ref{alg1} for treating the linearization error as uncertainty, can be given by replacing the constraints \eqref{admm constraint 2} in Algorithm \ref{alg1} with the constraints \eqref{modified admm constraints 2}.
\section{Simulation Results}
\label{Simulation Results}
%
\begin{figure}
\vspace{-0.2cm}
    \centering
\subfloat[ Trajectory with NTO \label{1a}]{
       \includegraphics[width= 0.455\columnwidth, trim={0.45cm 0.2cm 1.25cm 1cm},clip]{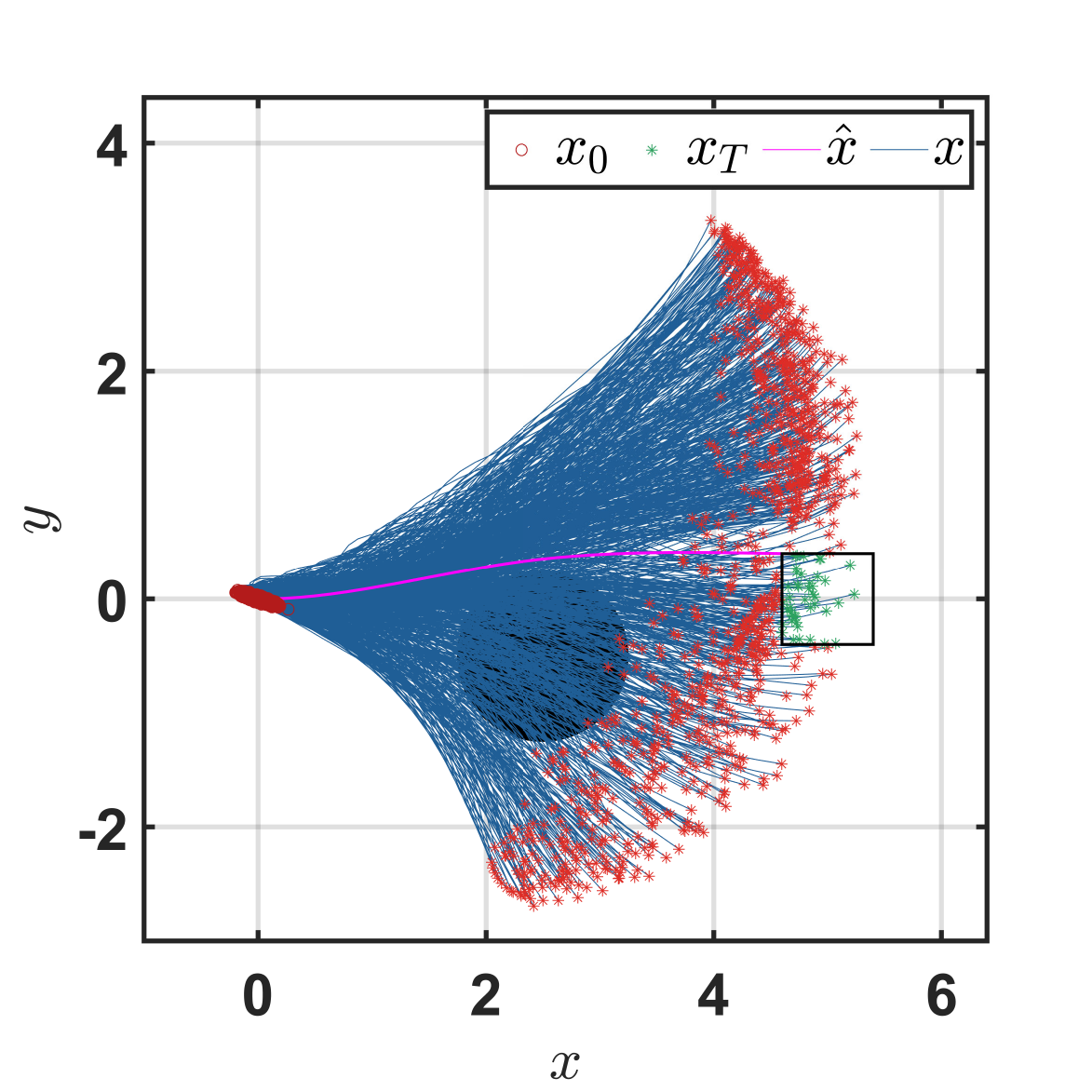}}
\subfloat[ Trajectory with NRTO \label{1b}]{
        \includegraphics[width=0.455\linewidth, trim={0.45cm 0.2cm 1.25cm 1cm},clip]{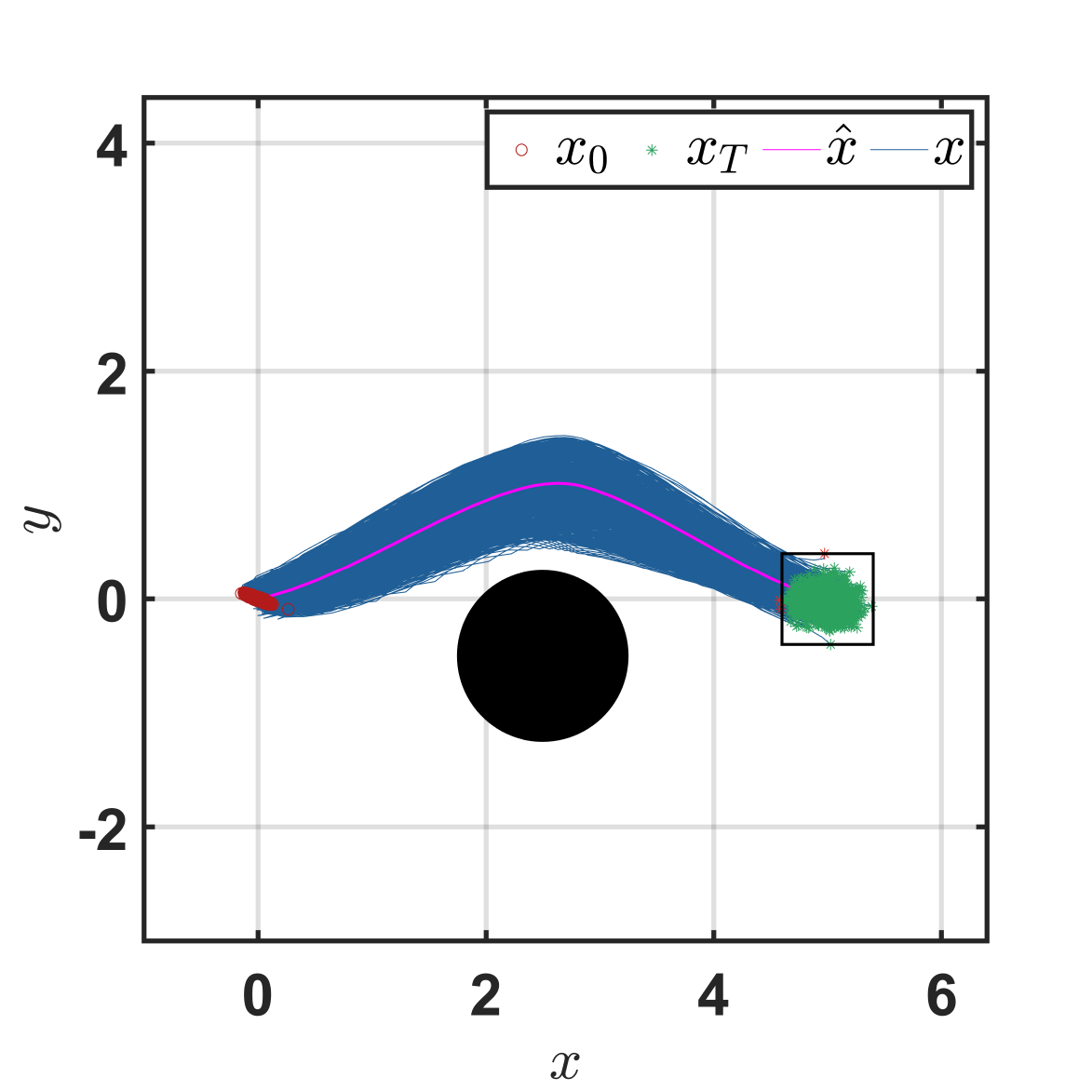}}
\\
  \caption{\textbf{NTO vs. NRTO:} 
  Only 2 out of 1000 trajectory realizations obtained using NTO \textit{satisfy} all the constraints. While only 3 out of 1000 trajectory realizations obtained using NRTO \textit{violate} the constraints.
  }
  \label{fig1} 
\end{figure}

In this section, we demonstrate the effectiveness of the proposed frameworks. Throughout this section, we refer to Algorithm \ref{alg1} without considering the linearization errors as \textbf{NRTO}, and to the extension for handling linearization errors as disclosed in Section \ref{Linearization error as uncertainty} as \textbf{NRTO-LE}. All simulations were carried out in Matlab2022b \cite{MATLAB} environment using YALMIP \cite{lofberg2004yalmip} as the modeling software and MOSEK \cite{mosek} as the solver on a system with an Intel Corei9-13900K.

We initially consider a unicycle model with state $[x; y; \theta]$, and control inputs $[v; \omega]$, where $(x,y)$ and $\theta$ represent the 2D position and angle respectively, while $v$ and $\omega$ represent the linear and angular velocities. The dynamics are provided in Appendix \ref{Simulation Dynamical models}. We consider a time horizon of $T=30$ with a step size $\Delta t = 0.01s$, and the uncertainty set parameters $\vS = \vI$, with $\vGamma$ and uncertainty level $\tau$ varying across the cases. The matrix $\vGamma$ is randomly generated in each case with each element $\Gamma_{ij} = \texttt{unifrnd}(-1,1)$. 
%


First, we demonstrate the effectiveness of our robust framework in Fig. \ref{fig1}.  We consider a non-robust variant of the framework NRTO, referred to as \textbf{NTO}, wherein the trajectory optimization problem is solved \textit{without} accounting for the disturbance $\bzeta$ (i.e., implementing NRTO with $\tau = 0$). For NRTO, we consider a case with an uncertainty level $\tau = 0.05$. The task requires the unicycle to reach the target position bounds (shown by the black box) while avoiding the circular obstacle. Fig. \ref{fig1} shows 1000 trajectory realizations obtained using each framework NTO and NRTO. 
Fig. \ref{1a} shows that using NTO, only a few reach the target position bounds (green markers inside the bounds). Further, only two realizations satisfy all the constraints (i.e., obstacle and terminal state constraints). On the other hand, using NRTO, as shown in Fig. \ref{1b}, all the trajectory realizations avoid the obstacle, and $99.7\%$ of them reach the target position bounds. Therefore, NRTO significantly improves constraint satisfaction over its non-robust variant (NTO), thereby enhancing the reliability of the system under uncertainty.

\begin{figure}[t!] 
    \centering
  \subfloat[ Trajectory \label{2a}]{
       \includegraphics[width= 0.51\columnwidth, trim={0.95cm 1cm 1.75cm 2.25cm},clip]{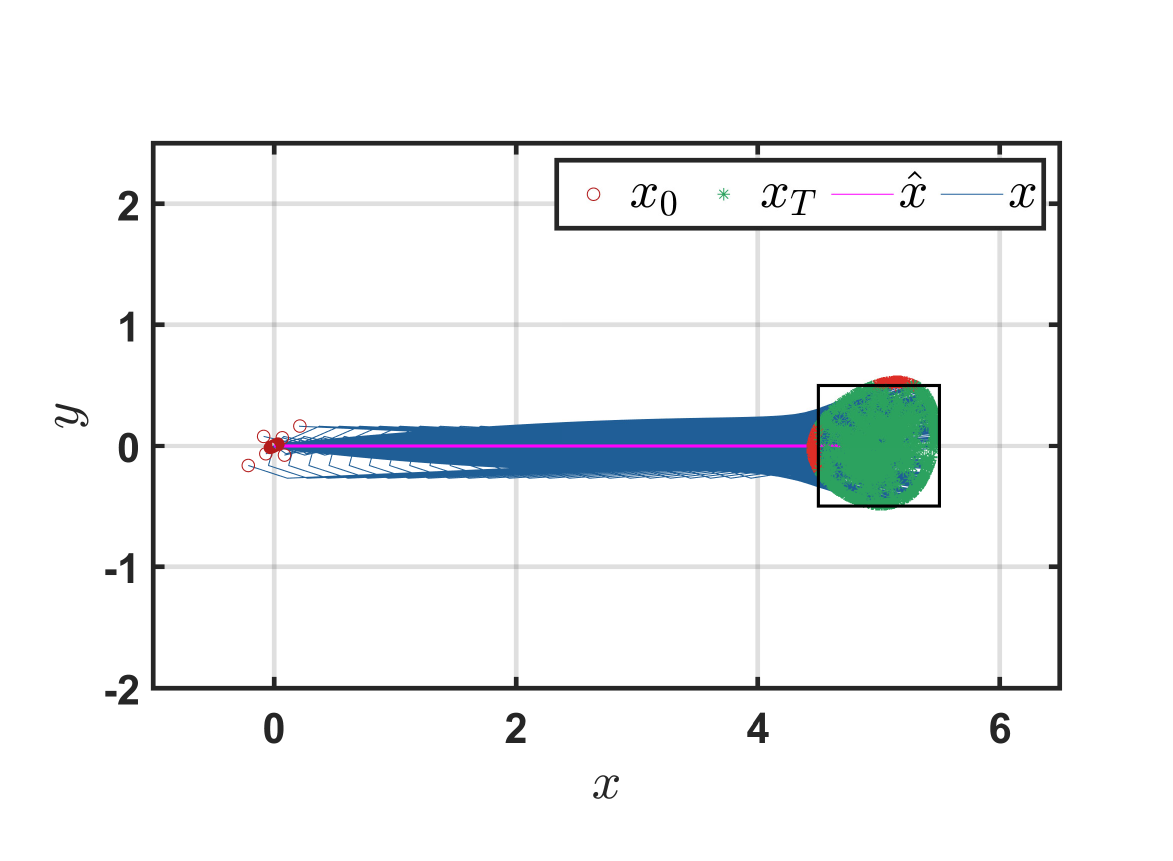}}
  \subfloat[ Terminal State  \label{2b}]{
        \includegraphics[width=0.36\linewidth, trim={0.2cm 0.10cm 0.3cm 1.25cm},clip]{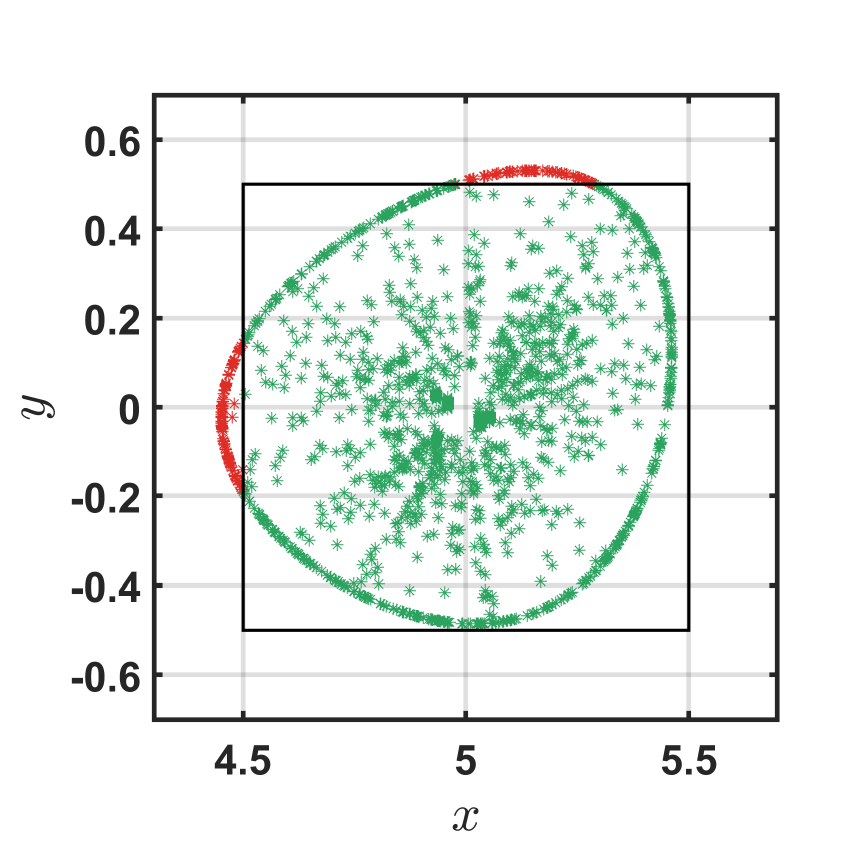}}
        \hfill
\subfloat[ Trajectory \label{2c}]{
        \includegraphics[width= 0.51\columnwidth, trim={0.95cm 1cm 1.75cm 2.25cm},clip]{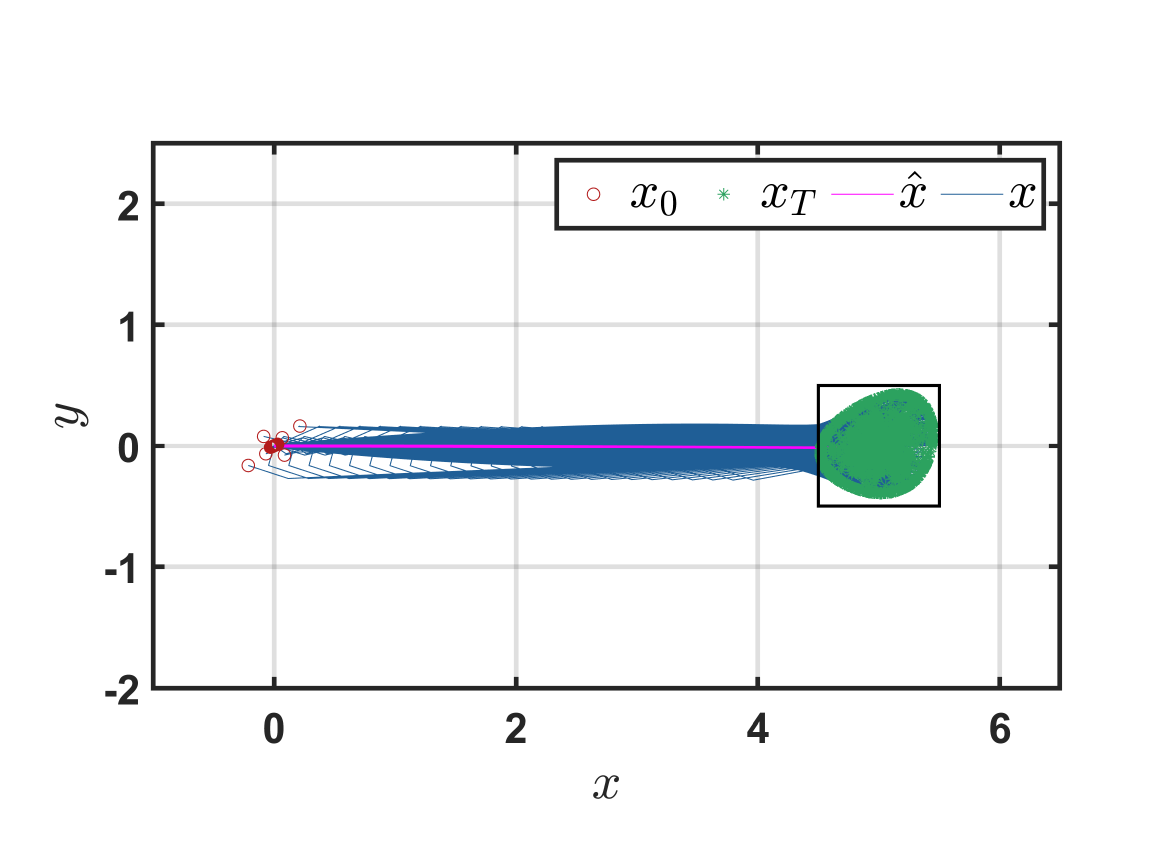} }
\subfloat[ Terminal State \label{2d}]{
        \includegraphics[width=0.36\linewidth, trim={0.2cm 0.1cm 0.3cm 1.25cm},clip]{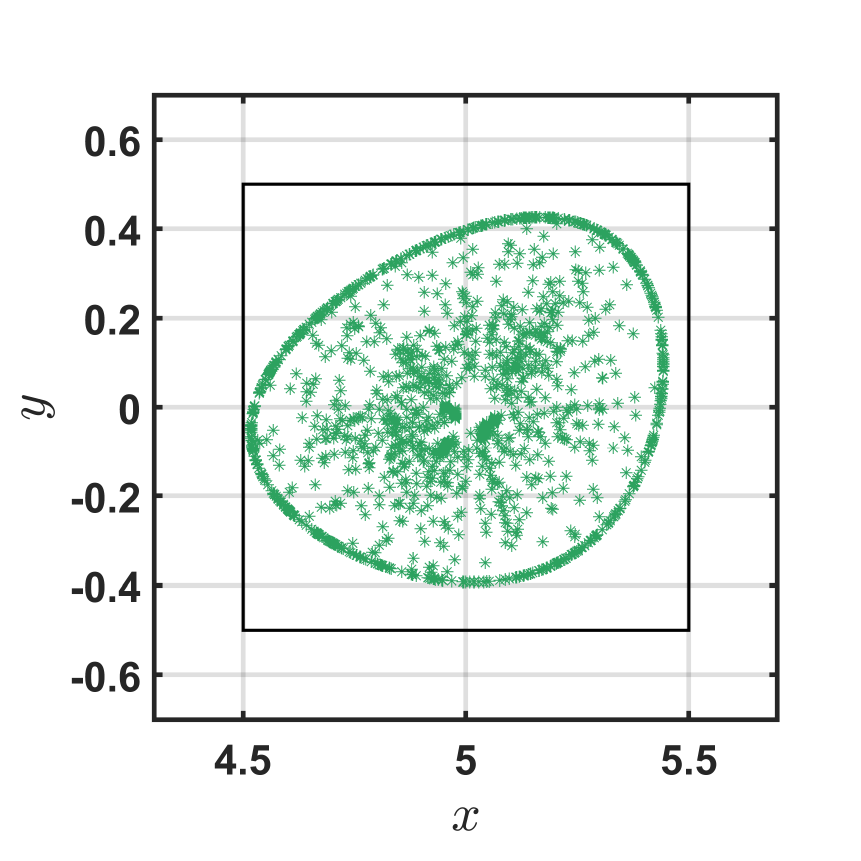}}
     \\
  \caption{ \textbf{Performance Comparison between NRTO and NRTO-LE:} Uncertainty level is set to $\tau = 0.1$. (a) and (b) correspond to the NRTO, with $92.1 \%$  constraint satisfaction. (c) and (d) correspond to the NRTO-LE, with $100 \%$ constraint satisfaction.}
  \label{fig2} 
\end{figure}
\begin{figure}[t!]
  \centering
       \includegraphics[width= 0.9\columnwidth, trim={0cm 0cm 0cm 0cm},clip]{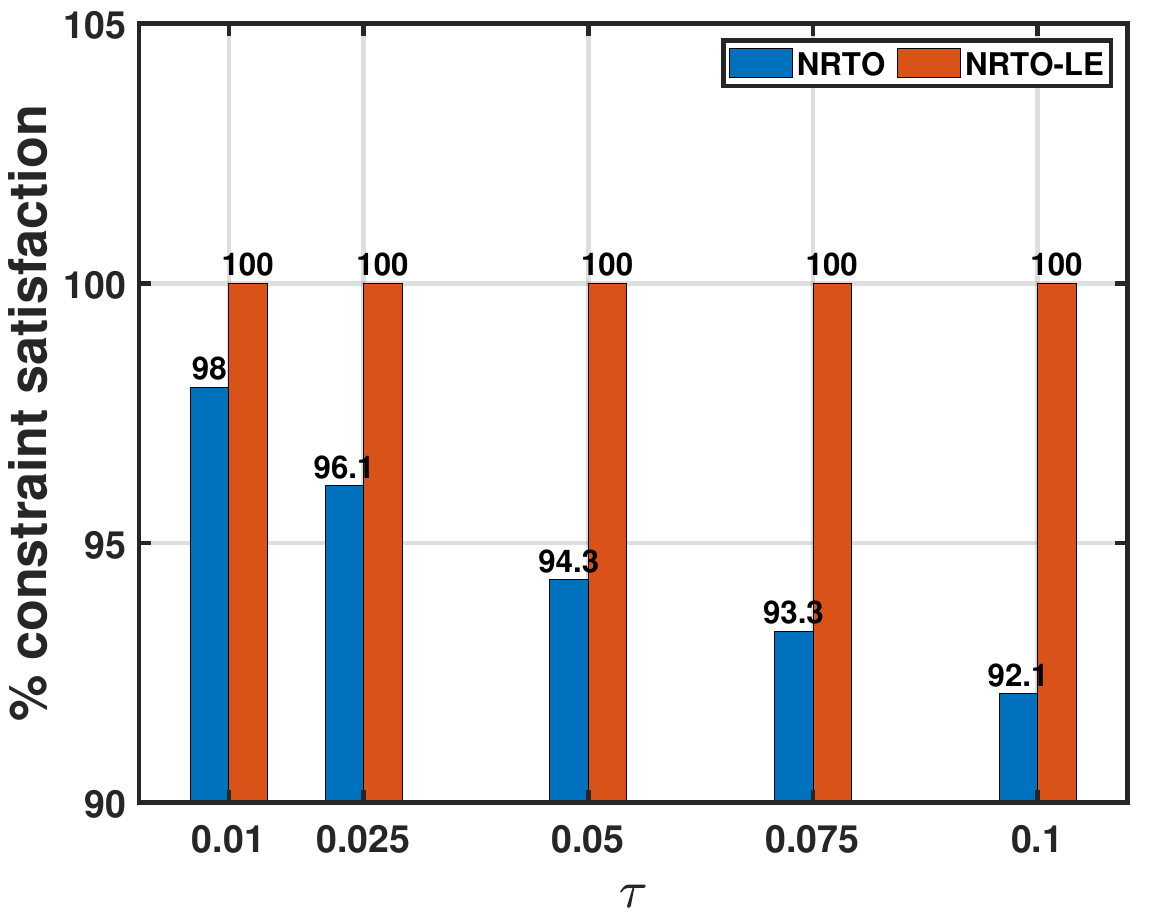} 
  \caption{\textbf{Uncertainty vs. Constraint Satisfaction:} Comparison of percentage of constraint satisfaction over 1500 realizations using NRTO and NRTO-LE. The constraint satisfaction decreases with increase in uncertainty using the NRTO, while NRTO-LE provides $100 \%$ constraint satisfaction in all the cases.  }
  \label{Fig3}
\end{figure}

As observed earlier in Fig. \ref{1b}, a few trajectory realizations obtained using NRTO still violate the terminal state constraints, which is due to the linearization error. Since the obstacle constraints involve concave functions - whose linearizations yield more conservative constraints - the effect of the linearization error is not that pronounced for these constraints. However, the effect can be observed for the terminal position constraints --- and it could be further amplified for models with stronger nonlinearities. In the following, we analyze the effect of the linearization error and highlight the effectiveness of NRTO-LE in guaranteeing robustness. In Fig. \ref{fig2}, we consider a case with an uncertainty level $\tau = 0.1$ and with the target position constraints as shown. Fig. \ref{2a} and \ref{2b} show the trajectory and terminal state realizations obtained using NRTO, while Fig. \ref{2c} and \ref{2d} show those obtained with NRTO-LE. Using NRTO only yields $92.1 \%$ constraint satisfaction, with a few terminal state realizations outside the target position bounds (red markings). To address this, we characterize the linearization error as disclosed in Section \ref{Linearization error as uncertainty} based on the trajectory realizations obtained using NRTO. In particular, we construct confidence ellipsoids from the aforementioned observed data to define the linearization error uncertainty sets of the form \eqref{lin const - error approach}.
Note that these linearization error uncertainty sets can also be constructed using other more sophisticated methods \cite{10528345}. 
Fig. \ref{2c} and \ref{2d} show that, using NRTO-LE,  none of the trajectory realizations violate the target position bounds, highlighting the robustness of the proposed approach.

\begin{figure}[t!] 
    \centering
  \subfloat[ Trajectory \label{4a}]{
       \includegraphics[width= 0.51\columnwidth, trim={0.35cm 0.25cm 1.75cm 1.15cm},clip]{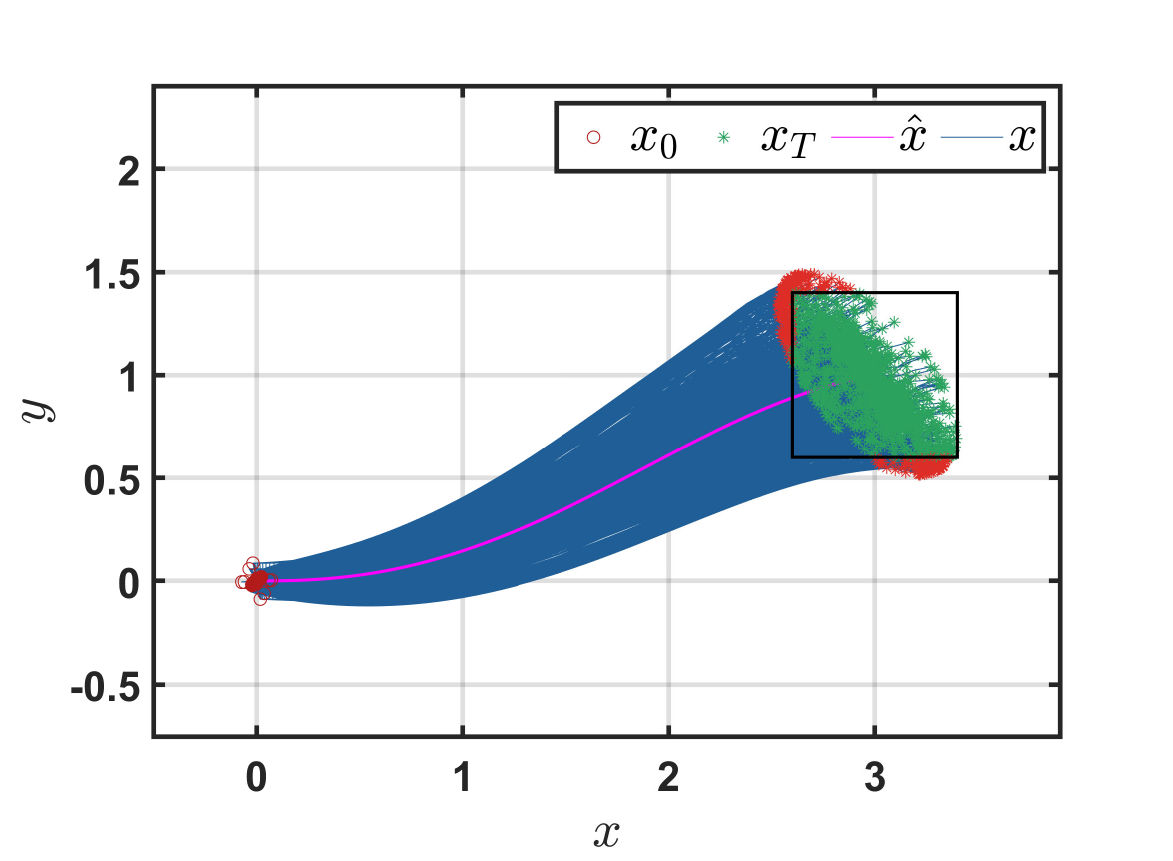}}
  \subfloat[ Terminal State  \label{4b}]{
        \includegraphics[width=0.365\linewidth, trim={0.35cm 0.25cm 1.25cm 1.15cm},clip]{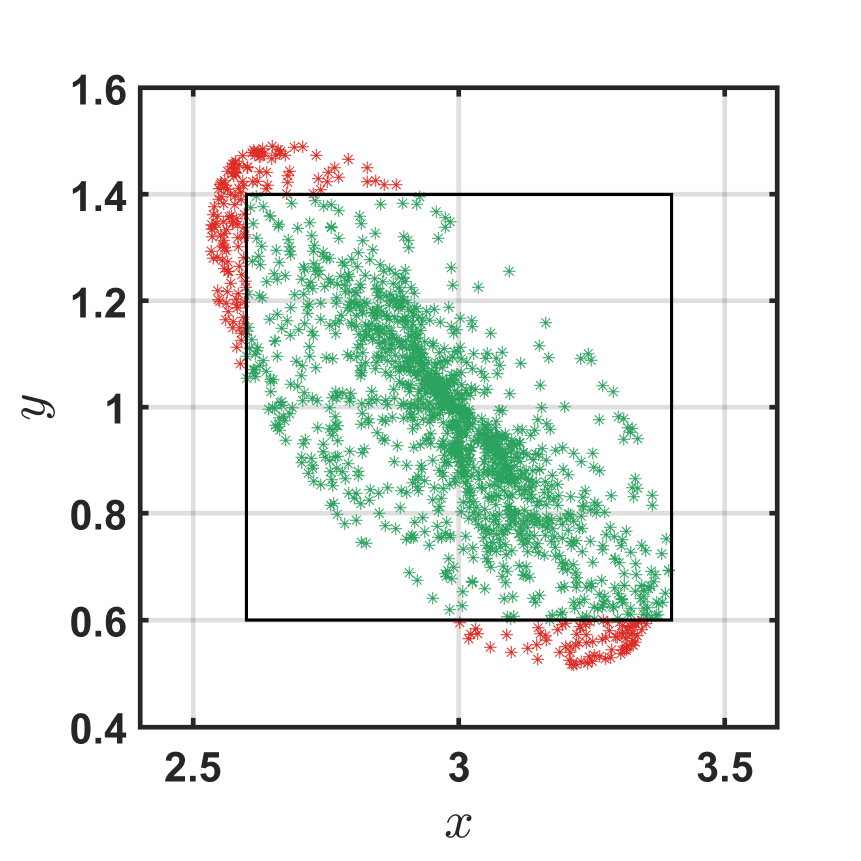}}
        \hfill
\subfloat[ Trajectory \label{4c}]{
        \includegraphics[width= 0.51\columnwidth, trim={0.35cm 0.25cm 1.75cm 1.15cm},clip]{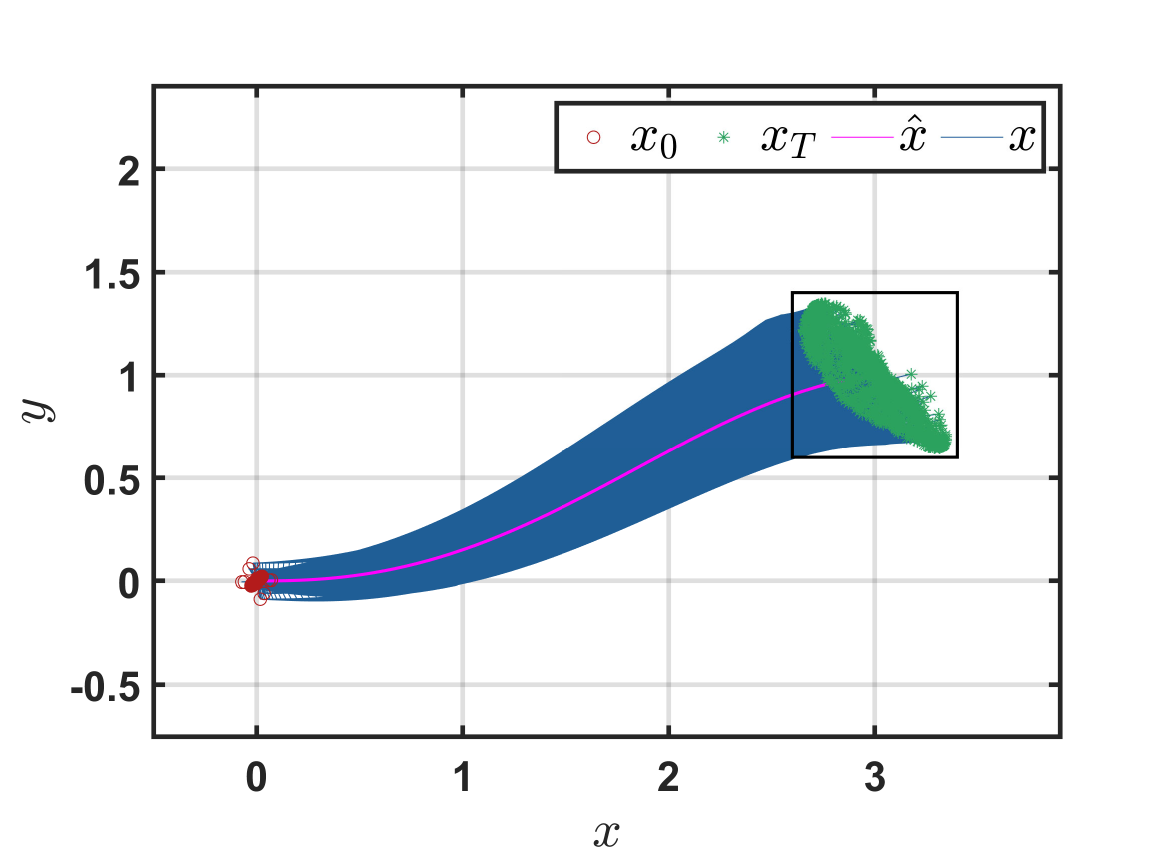} }
\subfloat[ Terminal State \label{4d}]{
       \includegraphics[width=0.365\linewidth, trim={0.35cm 0.25cm 1.25cm 1.15cm},clip]{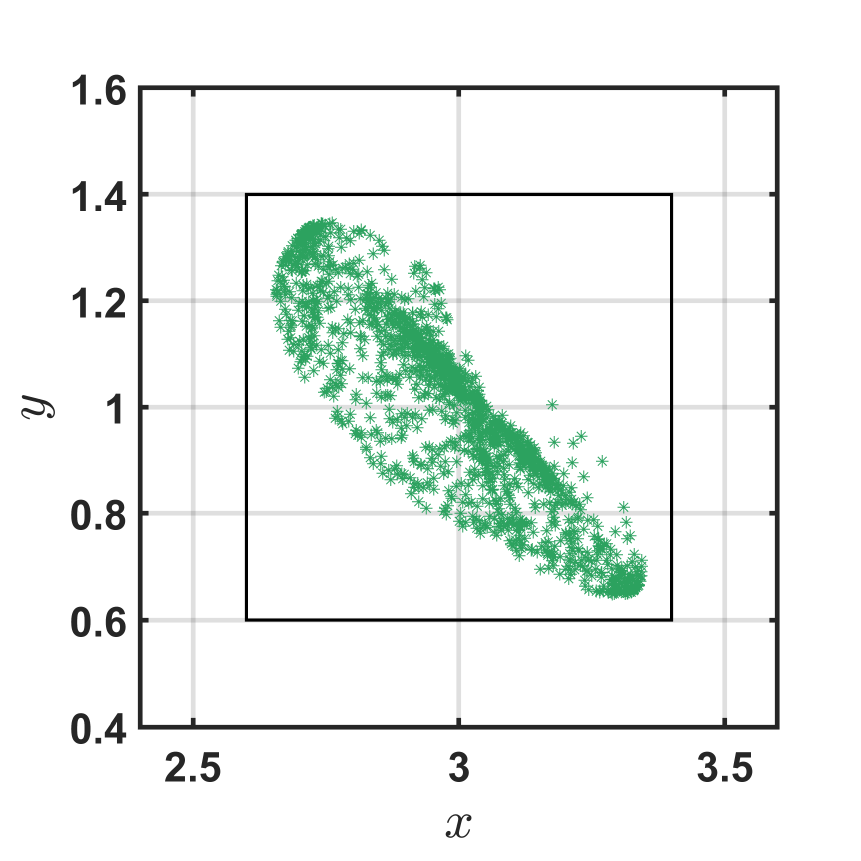} }
     \\
  \caption{ \textbf{Performance Comparison between NRTO and NRTO-LE for car model:} Uncertainty level is set to $\tau = 0.01$. (a) and (b) correspond to the NRTO, with $ 84.2 \%$  constraint satisfaction. (c) and (d) correspond to the NRTO-LE, with $100 \%$ constraint satisfaction. }
  \label{fig4} 
\end{figure}
%

\begin{figure}
\vspace{0.125cm}
  \centering
      \includegraphics[width= 0.95\columnwidth, trim={0cm 0cm 0cm 0cm},clip]{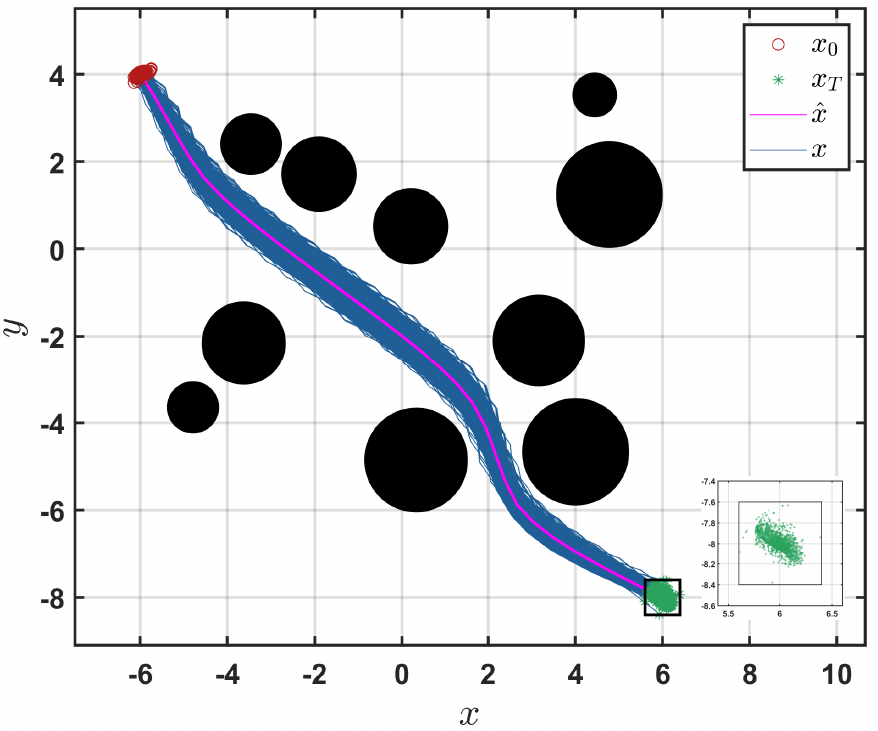} 
  \caption{\textbf{Complex scenario for unicycle with ten obstacles using NRTO-LE:} Uncertainty level is set to $\tau = 0.05$. $100 \%$ constraint satisfaction considering $1500$ trajectory realizations. }
  \label{Fig5}
\end{figure}

We further emphasize the effectiveness of NRTO-LE over NRTO by comparing their performance over increasing levels of uncertainty and nonlinearity. In Fig. \ref{Fig3}, we analyze the effect of the uncertainty level $\tau$ on the constraint satisfaction. This analysis considers the same task as in Fig. \ref{fig2} with the same $\vGamma$ and target position bounds. It can be observed that using NRTO, constraint satisfaction decreases with increasing levels of uncertainty. In contrast, NRTO-LE leverages the data obtained from the NRTO output and consistently provides $100 \%$ constraint satisfaction. In Fig. \ref{fig4}, we consider a more complex 2D car model with state $[x;y;\theta;v]$ and control inputs $[\omega; a]$. The state components $(x,y)$ represent the position of the midpoint of the back axle, $\theta$ represents the car's orientation, and $v$ represents the velocity of the front wheels. The control components $\omega$ and $a$ represent the angle and acceleration of the front wheels. The full dynamics are provided in Appendix \ref{Simulation Dynamical models}. We consider a time horizon $T= 50$ with a step size $\Delta t = 0.03s$.
The designated task requires the car to reach the target position bounds as shown in Fig. \ref{fig4}, and the terminal angle $\theta_T$ to be  constrained within $[-1,1]$. Fig. \ref{4a} and \ref{4b} correspond to NRTO, and Fig. \ref{4c} and \ref{4d} correspond to NRTO-LE. In both cases, there is no violation of the terminal angle constraint (i.e., $\theta_T \in [-1,1]$), yet, using NRTO, a considerable amount of realizations end up outside of the target position bounds --- with only $84.2 \%$ of them satisfying the constraints. On the other hand, the NRTO-LE approach provided $100 \%$ constraint satisfaction.

Lastly, we demonstrate the effectiveness of NRTO-LE in handling complex navigation tasks, as shown in Fig. \ref{Fig5}. This task involves a unicycle which is required to reach the target position bounds while avoiding ten circular obstacles under an uncertainty level of $\tau = 0.05$. Fig. \ref{Fig5} illustrates that none of the $1500$ realizations violate any of the constraints. 


%

%

%
%
%
\section{Conclusion}
\label{Conclusion}
We propose a novel, robust nonlinear trajectory optimization framework capable of handling nonlinear and nonconvex constraints. We leverage robust optimization (RO) techniques by effectively integrating them with sequential linearization schemes and an ADMM approach. Additionally, we effectively address the linearization error arising in such  schemes by modeling it as a deterministic disturbance. Our NRTO-LE framework demonstrates $100 \%$ constraint satisfaction in all studied cases and maintains its effectiveness for more complex scenarios.

In future work, we plan to extend the current framework for nonlinear robust MPC under unknown disturbances to handle dynamically evolving environments. We also aim to explore distributed optimization approaches for scaling our methodology to large-scale multi-agent systems \cite{abdul2024scaling}. Another promising direction would be investigating data-driven frameworks for estimating the uncertainty sets \cite{bertsimas2018data, wang2023learning} as well as for accelerating the underlying ADMM method \cite{saravanos2025deep}, towards further enhancing the robustness and scalability of the proposed approach. 
\appendix
\subsection{Linearized Dynamics matrices}
\label{Linearized Dynamics matrices}
The matrices $\vF_0$, $\vF_u $, $\vF_d$ and $\vF_{\zeta} $ are given as 
%
\begin{align}
    \vF_0
    & = 
    \begin{bmatrix}
    \vI ; & 
    \vPhi(1,0); & 
    \vPhi(2,0); &
    \vdots; &  
    \vPhi(N,0) 
    \end{bmatrix}, \nonumber
    \\
    \vF_u 
    & =
    \begin{bmatrix}
    \vzero & \vzero & \dots & \vzero \\ 
    \vB_0 & \vzero & \dots & \vzero \\ 
    \vPhi(2,1) \vB_0 & \vB_1 & \dots & \vzero \\ 
    \vdots & \vdots & \vdots & \vdots\\ 
    \vPhi(N,1) \vB_0 & \Phi(N,2) \vB_1 & \dots & \vB_{N-1}
    \end{bmatrix}, \nonumber
    \\
    \tilde{\vF}_{d} 
    & =
    \begin{bmatrix}
    \vzero & \vzero & \dots & \vzero \\ 
    \vI & \vzero & \dots & \vzero \\ 
    \vPhi(2,1) & \vI  & \dots & \vzero \\ 
    \vdots & \vdots & \vdots &  \vdots\\ 
    \vPhi(N,1) & \vPhi(N,2) & \dots & \vI
    \end{bmatrix}, \nonumber
    \\
    \vF_d & = \begin{bmatrix}
        \vzero & \tilde{\vF}_{d} 
    \end{bmatrix}, ~~
    \vF_{\bzeta} = \begin{bmatrix}
        \vF_0, \tilde{\vF}_{d} 
    \end{bmatrix}, \nonumber
\end{align}
where $\vPhi(k_1,k_2) = \vA_{k_1 -1} \vA_{k_1-2} \dots \vA_{k_2}$ for $ k_1 > k_2$.
\subsection{Simulation Dynamical models}
\label{Simulation Dynamical models}
Here, we provide the models used in our simulations. 
\paragraph{Unicycle Model} The unicycle dynamics are 
\begin{equation}
\begin{aligned}
    x_{k+1} & = x_k + v_k \cos(\theta_k) \Delta t,
    \\
    y_{k+1} & = y_k + v_k \sin(\theta_k) \Delta t,
    \\
    \theta_{k+1} & = \theta_k + \omega_k \Delta t.
\end{aligned}
\nonumber
\end{equation}
\paragraph{Car model} For the 2D car, the distance between the front and rear axles of car is $c_{\text{len}} = 0.75$. The rolling distances of the front and back wheels are $c^f_k = v_k \Delta t$ 
and
\begin{equation}
    c^b_k = c_k^f \cos(\omega_k) + c_{\text{len}} + \sqrt{c_{\text{len}}^2 - (c_k^f \cos(\omega_k))^2}, \nonumber
\end{equation}
respectively.
The dynamics are then given as
\begin{equation}
\begin{aligned}
    x_{k+1} & = x_k + c^b_k \cos(\theta_k), \quad y_{k+1} = y_k + c^b_k \sin(\theta_k),
    \\
    \theta_{k+1} & = \theta_k + \arcsin\bigg(\sin (\omega_k) \frac{c^f_k}{c_{\text{len}}} \bigg),
    \nonumber
    \\
    v_{k+1} & = v_k + a_k \Delta t.
\end{aligned}
\end{equation}

\bibliographystyle{IEEEtran}
\bibliography{IEEEabrv, references}

\end{document}